\documentclass[12pt]{article}
\begin{document}
\begin{center}
{\bf
Quark-Hadron Duality in Photoabsorption Sum Rules and
Two Photon Decays of Meson Resonances
}
\vskip 5mm

S.B. Gerasimov$^{ \dag}$
\vskip 5mm
{\small
{\it
Bogoliubov Laboratory of Theoretical Physics\\
Joint Institute for Nuclear Research,
141980 Dubna (Moscow region), Russia}
\\
$\dag$ {\it
E-mail: gerasb@thsun1.jinr.ru
}}
\end{center}
\vskip 5mm
\begin{center}
\begin{minipage}{150mm}
\centerline{\bf Abstract}
The idea of quark-hadron duality is developed and applied to
integral sum rules for the photoexcitation of meson resonances.
Some applications of the presented  approach in the
light and heavy quark sectors are made, and the role of the
scalar diquark cluster degrees of freedom
in the radiative formation of light scalar mesons is discussed.
\end{minipage}
\end{center}
\vskip 10mm

\section{Introduction }
In the present report, we continue consideration of aspects
of the quark hadron duality, {\it i.e.} the equivalence of
two complete sets of state vectors, saturating certain integral
sum rules, one of the sets being the solution of the bound state problem
with colour-confining  interaction, while the other describes free partons.
Further exploration of relations between exclusive and inclusive production of hadrons in
processes induced by virtual and real photons is worthwhile as means
of more extensive tests of approximations accompanying the description
of the transition between
hadron and quark-gluon degrees of freedom in QCD.
The choice of sum rules satisfying the assumed duality
condition is suggested by correspondence with the well-known results in the
nonrelativistic theory of interaction of the radiation with matter.
The
message is that sum rules connected with the dipole moment fluctuation seem
to be singled in both nonrelativistic
\cite{KhFo}
and relativistic regions
\cite{Ge69,Ge75,Ge79}.
Following this idea, which has first been tested
in the models of quantum  field theory,
we present, within the relativistic constituent quark model approach,
the relations between
two-photon decay widths of the lowest meson resonances following from
the derived sum rules
for  polarized gamma-gamma cross sections.
\section{
Quark-hadron duality for
the bremsstrahlung - \\
-weighted sum rules. The case of meson resonance
photoexcitation
}

We first remind that applying the adopted duality principle
of two complete sets of final state vectors (one - with the confined
$q\bar q$-states representing the sum over all resonances, the other-
with the "gedanken" free $q\bar q$ -pair) to the Cabibbo-Radicati
sum rule
\cite{CR} for the pion, we get
\cite{Ge79}
\begin{eqnarray}
<r^2>_{\pi^{\pm}}={\frac{3}{4\pi^2 \alpha}}\int_0^\infty
\frac{d\nu}{\nu}(\sigma(\gamma^{-} \pi^{+}) - \sigma(\gamma^{+} \pi^{+}))
= \frac{3}{4\pi^2 F_{\pi}^2}
\end{eqnarray}
Further useful relations were obtained
\cite{Ge79}
from the assumed approximate equality
of the derivatives
$$F^{\prime}_{\gamma^{*}\gamma \pi^{o}}(q^2)|_{q^2 \to 0} \simeq
F^{\prime}_{\gamma^{*} \pi^{\pm} \pi^{\pm}}(q^2)|_{q^2 \to 0}$$
of form factors normalized
to unity at zero momentum transfers
\begin{eqnarray}
m_q & \simeq &
\sqrt{\frac{2}{3}} \pi F_{\pi},(q=u,d)
\\
\frac{g_{\pi^{o}qq}^2}{4\pi} &
\simeq & \frac{\pi}{6}
\end{eqnarray}
where the numerical value of the
pseudoscalar $\pi qq $-coupling constant followes from an analogue of
the Goldberger-Treiman relation for quarks.

The successful local approximation for the $\pi q\bar q$~-vertex
in the calculation
of $<r^2>_{\pi}$ means that it should be approximately valid in
computation of the radiative decays $M^{*}\to \gamma \pi$~~of
the lowest meson resonances contributing to the sum rule (1).
Therefore, a similar mechanism of the local annihilation of
constituent quarks in the reactions $q\bar q \to e^{+}e^{-}$~~ and
$q\bar q \to \gamma \pi$~~ enables one to obtain, {\it e.g.},
for the $\omega$~-meson
\cite{Ge79}
\begin{eqnarray}
\frac{\Gamma(\omega \to  e^{+}e^{-})}{\Gamma(\omega \to \pi^{0}\gamma)}\,
\simeq \frac{\sigma((q\bar q)_{\omega} \to e^{+}e^{-})}
{\sigma((q\bar q)_{\omega} \to  \pi^{0} \gamma))}|_{s=m_{\omega}^2}
\simeq \frac{\alpha}{18}\,(\frac{g_{\pi q\bar q} ^2}{4\pi})^{-1}=\frac {\alpha}{3\pi}
\end{eqnarray}
where $(q\bar q)|_{\omega}=(1/{\sqrt{2}})(u\bar u+d\bar d)$
and numerically the ratio of the widths
turns out to be within the experimental uncertainties of the data.
The seemingly paradoxical relevance of the local approximation
combined with the application of the Goldberger-Treiman relation
to the pion-quark vertex as compared to an appearing more general "softened"
description in which this vertex is described by a Bethe-Salpeter type
equation, may be interpreted as an exemplification, in the considered
context, of the twofold picture of the pion.
As is well-known, the pion presents itself in twofold ways:
on the one hand, its anomalously small mass suggests that it should be
identified with the pseudo-Goldstone mode of dynamical breaking of chiral
symmetry in QCD; on the other, it should be described equally well from
the QCD Lagrangian in terms of current quarks,
or in terms of the QCD motivated relativistic models of
constituent quarks with strong
attractive forces acting in the $J^P=0^-$ channel.
The complementarity of these pictures may also reflect
a kind of duality between the effective hadronic description based on
symmetries, and the microscopic description in terms of partons.
This duality can be alternatively exploited in calculations
of hadron properties.
While the high momentum transfer reactions reveal characteristics
of the quark and gluon
distributions inside the pion, at low energies its role as a (pseudo)Goldstone
boson mode became essential in describing the long-range structure and
strength of its interaction with hadrons and effective hadronic constituent
degrees of freedom , e.g.,constituent quarks.
However, the inclusion of hadronic corrections to $\langle r^2\rangle_{\pi}$
in the form of the one-pion-exchange graph contribution to the
Cabibbo-Radicati sum rule, which provides the chiral corrections of the
order $\sim$~ log($m_{\pi})$
\cite{Ge75,Ge79}
in accordance with
the theorem of Beg and Zepeda
\cite{Beg},
definitely improves the
agreement with data and one should have in mind
this type of correction in other applications of the approach outlined.\\
We turn now to sum rules for meson
resonances in photon-photon collisions. Varying the polarizations of
colliding photons, one can show that the linear combination of certain
$\gamma \gamma \to q\bar q $ cross-sections will dominantly collect,
at low and medium energies most important for saturation of
the integral sum rules considered,
the $q\bar q$- states with definite spin-parity and hence, by the adopted
quark-hadron duality, the meson resonances with the same quantum numbers.
The polarization structure of the transition matrix element
$M(J^{PC} \leftrightarrow 2\gamma)$ for the meson resonance with the spin-parity
$J^{PC} = 0^{-+}, 0^{++}, 2^{\pm +}$ is taken as follows:
\begin{eqnarray}
M(0^{++}\leftrightarrow 2\gamma)&=&G_{\mathrm{S}}[(\epsilon_1 \epsilon_2)(k_1 k_2 ) -
(\epsilon_1 k_2)(\epsilon_2 k_1)],\\
M(2^{++}\leftrightarrow 2\gamma)&=&G_{\mathrm{T}0}[(\epsilon_1 \epsilon_2)
(k_1 k_2 )-(\epsilon_1 k_2)(\epsilon_2 k_1)]k_1^{\mu}k_2^{\nu}
{e^{\mu \nu}}\nonumber \\
&+&G_{\mathrm{T2}}[(k_1 k_2 )\epsilon_1^{\mu} \epsilon_2^{\nu} -
(\epsilon_1 k_2)k_1^{\mu}\epsilon_2^{\nu}]{e^{\mu \nu}} \nonumber \\
&-&G_{\mathrm{T2}}[(\epsilon_2 k_1)k_2^{\mu}\epsilon_1^{\nu}+
(\epsilon_1 \epsilon_2)k_1^{\mu}k_2^{\nu}]{e^{\mu \nu}},\\
M(0^{-+}\leftrightarrow 2\gamma)&=&G_{\mathrm{PS}}\varepsilon_{\mu \nu \lambda \sigma}
k_1^{\mu}k_2^{\nu}\epsilon_1^{\lambda}\epsilon_2^{\sigma},\\
M(2^{-+} \leftrightarrow 2\gamma)&=&G_{\mathrm{PT}}\varepsilon_{\mu \nu \lambda \sigma}
k_1^{\mu}k_2^{\nu}\epsilon_1^{\lambda}\epsilon_2^{\sigma}k_1^{\alpha}k_2^{\beta}
{e^{\alpha \beta}},
\end{eqnarray}
where $k_{i}^\mu$ are the momenta of photons,~$\epsilon_{i}^{\nu}~
({e^{\alpha \beta}})$
-polarization
vector (tensor) of the photon (the tensor meson), $G$ - corresponding
coupling constants, $G_{\mathrm{T\lambda}}$ being the tensor meson
coupling constants with the $z$-projection of the total angular momentum
$\lambda = 0$~~or~~$2$, respectively.

It follows then that the combinations of the integrals over the
bremsstrahlung-weighted and polarized
$\gamma \gamma \to q\bar q $ cross-sections,
$I_{\bot} - (1/2)I_{p},~~ I_{\|} - (1/2)I_{p},~~ I_{p}$ will be related to
low-mass meson resonances having spatial quantum numbers $J^{PC} =
0^{-+}$~ and~$ 2^{-+}$,
~~$0^{++}$~ and~$2^{++}(\lambda=0)$,
~~$2^{++}(\lambda=2)$, if we confine ourselves to the mesons with
spins $J\leq 2$ for further discussion.

The $\gamma \gamma $ - cross-sections $\sigma_{\bot (\|)}$
(and the integrals thereof) refer to
plane-polarized photons with the perpendicular (parallel)
polarizations, and $\sigma_{p} $  corresponds to
circularly polarized photons with parallel spins.

Evaluating cross-sections and elementary integrals
we get the sum rules for radiative widths of resonances with
different values of $J^{PC}$
\begin{eqnarray}
\sum_{i} \frac{\Gamma(PS_{i} \to 2\gamma)}{{m_{PS_{i}}^3}}
+ 5\sum_{j} \frac{\Gamma(PT_{j} \to 2\gamma)}{{m_{PT_{j}}^3}}\simeq
\frac{3}{16\pi^2} \sum_{q}{\langle Q(q)^2 \rangle}^2 \frac{\pi \alpha^2}
{m_{q}^2}
\end{eqnarray}
\begin{eqnarray}
\sum_{i} \frac{\Gamma(S_{i} \to 2\gamma)}{{m_{S_{i}}^3}}
+ 5\sum_{j} \frac{\Gamma((T0)_{j} \to 2\gamma)}{{m_{T0_{j}}^3}} \simeq
\nonumber \\
\simeq \frac{3}{16\pi^2}[ \sum_{q}{\langle Q(q)^2 \rangle}^2
\frac{5\pi \alpha^2}{9m_{q}^2} + \sum_{qq}{\langle Q(qq)^2 \rangle}^2
\frac{2\pi \alpha^2}{9m_{qq}^2}]
\end{eqnarray}
\begin{eqnarray}
5\sum_{i} \frac{\Gamma((T2)_{i} \to 2\gamma)}{{m_{T2_{i}}^3}}
+ (2J+1)\sum_{R(J\geq 2;\lambda=2)} \frac{\Gamma(R \to 2\gamma)}{m_{R}^3} \simeq
\nonumber \\
\simeq \frac{3}{16\pi^2} \sum_{q}{\langle Q(q)^2 \rangle}^2 \frac{14\pi
\alpha^2} {9m_{q}^2}
\end{eqnarray}
All the integrals over the resonance
cross sections are taken in the narrow width approximation
and, further, the contributions of the states with $J\geq 3$
will be neglected.
For generality of the consideration,
we included the last term in (10) that corresponds to a possible role of
scalar diquarks as constituent "partons" composing, at least in part,
the scalar meson nonets
(for discussion of this acute problem in hadron spectroscopy
see, {\it e.g.}, minireview in
\cite{PDG}
and references therein).
Assuming no mixing between light and heavy quark sectors, one can read every
sum rule separately  for mesons constructed of the u-,d-, s-, and
c-quarks. Moreover, one can split sum rules into the isovector and isoscalar
resonance parts in the light quark sector, which enables one to carry a
more detailed comparison with experimental data.
For numerical estimation of radiative widths we use the mass values
$m_c=1.5$~GeV, $m_u=m_d \simeq 240$~MeV, $m_s$ is taken from \cite{Ge79}
as expressed via the strangeness-changing GT-relation, $m_s \simeq 350$~MeV,
and for the diquark masses we take the values following from
the extended chiral model including scalar diquarks into the
low-energy effective Lagrangian approach
\cite{Pr99}:
$m_{qq} \simeq 310 \div 330$~MeV, $m_{qs} \simeq 545 \div 570$~MeV, where
$q=u,d$.

In the charm sector, from our sum rules we have obtained
$\Gamma_{\gamma \gamma}(\eta_c) \simeq 7.4~(6.9\pm 1.7 \pm 0.8)$,
$\Gamma_{\gamma \gamma} (\chi_{c0}) \leq 6.1~(3.76\pm 0.65\pm 1.81)$,
$\Gamma_{\gamma \gamma} (\chi_{c2}) \leq 3.9~[(0.53\pm 0.15\pm 0.23)] \div
[(1.76\pm 0.47\pm 0.37)] $ where the most recent results on the
two-photon widths of resonances in the units $keV$ from
\cite{Bra00}
are given in parentheses and a smaller (larger) value for
the width $\Gamma_{\gamma \gamma} (\chi_{c2})$ refers to the registered
$\chi_{c2}$~- decay final states $2\pi^{+}2\pi^{-}$
($l^{+}l^{-}\gamma$) . The closeness of
$\Gamma_{\gamma \gamma} ^{exp}(\eta_c)$
to the value calculated from sum rules points to relative smallness
of the higher resonance part of the spectrum, {\it e.g.}, the radial
excitations of pseudoscalar charmonia compared
to the ground state contribution.
The situation with the scalar charmonium seems to signal about
a more important role of higher radial excitations.
The cross section of the transition
$\gamma \gamma \to q\bar q(J^{PC}=2^{++})$
is known
\cite{Rob86}
to fall much slower with rising photon
energies as compared to the transitions to the states with $J^{PC}=0^{\pm
+}$, and this is reflected by rather a large contribution from the higher
resonances with larger masses and spins.
The situation with the light tensor mesons is largely the same,
and we shall instead focus on the pseudoscalar and scalar meson sum rules
in the light quark sector.

Remarkably enough, the small upper bound of
the radiative width of the pseudotensor
$\pi_{2}(1670)$-resonance, reported recently by the L3
and ARGUS Collaborations
\cite{L3-97,ARGUS} together with the known two-photon widths of
$\pi^0$-, $\eta$-, and $\eta^{\prime}$-mesons
provide the fulfilment of the pseudoscalar sum rule within experimental
uncertainties.

Concerning the nature of the low-lying scalar mesons,
there is still no general agreement on
where are the  $q\bar q$ states, whether there is
a glueball among the light scalars, and whether some
of the many scalars are multiquark,
or meson-meson bound states.
One of the most economic viewpoints would be to try to identify low-mass
scalars with the ground and radial excited $q\bar q$ states
and the glueball mixed with quarkonia-type states
\cite{Anis02,Vol}.
Another popular scheme consists in the hypothesis that
above 1 GeV the scalar states form a
conventional $q\bar{q}$ nonet mixed with the  glueball,
suggested by lattice QCD,
while below 1 GeV the states also form a nonet,
but of a more complicated nature
\cite{Ach98,Black98,ClT02}.
Namely, as implied by the attractive forces
of QCD, the diquark configurations of the type
 $(qq)_{\bar{3}}(\bar{q}\bar{q})_3$ in S-wave,
possibly, with some $q\bar{q}$ admixture
in P-wave, can form a number of broad and low-lying resonances,
which can also be identified with observed states.
We apply the sum rule (12) for light scalar mesons
to test both the afore-mentioned pictures of two nonets of scalar mesons.
For the isovector mesons $a_{{S_1}} \equiv a_{0}(980)$ and
$a_{{S_2}} \equiv a_{0}(1450)$~~ we have
\begin{eqnarray}
\sum_{i=1,2}\frac{\Gamma_{\gamma \gamma}(S_{i})}{m(S_{i})^3} \simeq
\frac{\alpha^2}{96\pi}(\frac{5}{9{m_{q}}^{2}}+\frac{2}{9{m_{qs}}^2})
\end{eqnarray}
We take $\Gamma_{\gamma \gamma}(a_0(980)) \simeq 0.3$~keV~
\cite{PDG},
and $m_{q,(qs)}= 240 (550)$~MeV~ for a further numerical calculation.
Let us consider first the "ground + radial-excited" $q\bar q$-option
for $a_0(980)$ and $a_0(1450)$-resonances. It follows then from
the sum rule (12) with the second term ({\it i.e.}, the diquark term,
containing $m_{qs}$) omitted in the right-hand side:
\begin{eqnarray}
\Gamma_{\gamma \gamma}(a_0(1450))&\simeq& 2.9~~ keV\nonumber \\
\frac{\Gamma_{\gamma \gamma}(S_{1})}{m(S_{1})^3}:\frac{\Gamma_{\gamma \gamma}(S_{2})}{m(S_{2})^3}&\simeq& 1:4.34
\end{eqnarray}
While the absolute value of  $\Gamma_{\gamma \gamma}(a_0(1450))$
looks reasonable,
the ratio of the "reduced couplings" squared (the second line in Eq.(13))
is not, in our opinion. In the potential
nonrelativistic models, this ratio can essentially be
interpreted as a ratio of derivatives of the meson radial
wave function at the "zero" interquark distance:
$|R_{{S_1}}^{\prime}(0)|^2:|R_{{S_2}}^{\prime}(0)|^2$,
and the situation of the ground-to-(1st)radial-state ratio such as
given by Eq.(13) looks disfavouring (for example, in heavy quark systems
obeying the Schr\"odinger equation with the QCD-motivated potentials
$|R_{1P}^{\prime}(0)|^2:|R_{2P}^{\prime}(0)|^2\simeq 1:1.2$~
for the nP-wave states
\cite{PTN86}).

Following another option, we identify $a_0(980)$ and $a_0(1450)$ as
 generally mixed configurations $(1/\sqrt{2})(u\bar{u}-d\bar{d})$
in P-wave and
$(1/\sqrt{2})((us)(\bar{u}\bar{s})-(ds)(\bar{d}\bar{s}))$ in S-wave.
First, we note that if one takes $a_0(980)$ as the pure $2q2\bar{q}$ and
$a_0(1450)$ as the pure $q\bar{q}$~-configuration, then one can obtain
from the sum rule (12)~~$\Gamma_{\gamma \gamma}(a_0(980)) \leq 0.12$~keV~
and $\Gamma_{\gamma \gamma}(a_0(1450)) \leq 5.2$~keV.
The two-photon decay width of $a_{0}(980)$ turns out essentially lower
than accepted value $0.3$~keV in that case. Therefore, it appears
reasonable to correct the situation with the help of the mixing
of two parton configurations. Tentatively, we accept the simplest
case of the orthogonal mixing of two
isovector $|q\bar q>$ and $|(qs)(\bar q \bar s)>$ basis states,
assuming also $\Gamma_{\gamma \gamma}(a_0(980))=0.3$~keV.
Irrespectively of the mixing, the value
$\Gamma_{\gamma \gamma}(a_0(1450)) \simeq 3.5$~keV follows just from the sum rule.
Solving the following equation:
\begin{eqnarray}
\frac{\Gamma_{\gamma \gamma}(S_1)}{m(S_1)^3} &:&
\frac{\Gamma_{\gamma \gamma}(S_2)}{m(S_2)^3} = \nonumber \\
{[\frac{\sqrt{2}}{3m_{qs}}cos{\theta} +
\frac{\sqrt{5}}{3m_{q}}sin{\theta}]}^{2} &:&
{[-\frac{\sqrt{2}}{3m_{qs}}sin{\theta}+
\frac{\sqrt{5}}{3m_{q}}cos{\theta}]}^{2},
\end{eqnarray}
for the mixing angle $\theta$, one obtains two solutions:
$\theta \simeq 12^{o}$ and $\theta \simeq 137^{o}$, where the first (second)
corresponds to the positive (negative) relative sign of the amplitudes
$A(a_{0}(980)\to 2\gamma)$ and $A(a_{0}(1450)\to 2\gamma)$.
The quark-flavour structure of physical resonance states is then
\begin{eqnarray}
a_0(980)=0.691((us)(\bar{u}\bar{s})-(ds)(\bar{d}\bar{s}))+0.147(u\bar{u}-d\bar{d})\nonumber \\
a_0(1450)=- 0.147((us)(\bar{u}\bar{s})-(ds)(\bar{d}\bar{s}))+0.691(u\bar{u}-d\bar{d}),
\end{eqnarray}
for the positive sign, and
\begin{eqnarray}
a_0(980)\simeq \frac{1}{2}(-(us)(\bar{u}\bar{s})+(ds)(\bar{d}\bar{s}))+u\bar{u}-d\bar{d})\nonumber \\
a_0(1450)\simeq -\frac{1}{2}((us)(\bar{u}\bar{s})-(ds)(\bar{d}\bar{s})+u\bar{u}-d\bar{d}),
\end{eqnarray}
for the negative one.

A smaller value of the mixing angle $\theta$ seems to conform better
to the analysis of the strong decays of light scalars\cite{Black98}.
The predicted large radiative width $\Gamma_{\gamma \gamma}(a_{0}(1450)$
may present a potential difficulty for the explanation of the apparent
absence of a resonance effect due to the excitation of $a_0 (1450)$
in the mass spectrum of the ${K_{S}}^0{K_{S}}^0$- pairs
produced in the $\gamma \gamma$~-reaction
\cite{L3-01}.
However, the detailed
analysis of this situation requires simultaneous consideration of
excitation of both isovector and closely lying isoscalar scalar resonances,
which are expected (by analogy with the known effect of the tensor $a_2(1320)$~-
and $f_2(1270)$~- resonance interference)
to interfere destructively in the $K^{0}{\bar K}^{0}$  decay channel.
Unfortunately, it is presently impossible to analyse,
via analogous sum rules, the very interesting and important isoscalar
sector of light scalar mesons due to the lack of needed data on radiative widths.
\section{
Concluding remarks
}
We believe that the sum rules for resonance $\gamma \gamma$ interaction
give, on the average, a good evidence for the relevance of quark- hadron
duality in the considered context.

1.The  application of the developed approach
to derive and check the $\gamma \gamma$- sum rules
leads to especially good results for pseudoscalar mesons
in both light and charm quark sector. A rapid decrease in  the corresponding
polarized $\gamma \gamma$~-cross section in the integral sum rule
explains the sum rule saturation by the ground states of pseudoscalar mesons.\\
2.As it follows from sum rules to be saturated by the light scalar mesons,
the $\gamma \gamma$~- resonance couplings can give important information
on the flavour and parton content of these states. One can hope that
such information on the scalar mesons
will be an essential part of interpreting these states.
Further, the $Q^2$ dependence of the reaction
$\gamma^*\gamma \to  f_0/a_0$ on the photon virtuality
could probe the spatial dependence of the wave function of these states.\\
3.The measurements of radiative decays
of higher spin meson resonances would be very desirable and interesting
in view of a demonstrated stronger dependence of the sum rule, including
the higher spin, $J\geq 2$, on higher mass intermediate states
and photon energies.\\
\noindent
In conclusion, the author wishes to thank the organizers
of the GDH-2002 Conference for invitation, warm hospitality, and support.\\
\noindent
This work was partially supported by the RFBR grant No. 00-15-96737.

\end{document}